\documentclass[pra,10pt,notitlepage,amsmath,twocolumn,superscriptaddress]{revtex4-2}
\usepackage{graphicx}
\begin{document}

\title{Nonclassical correlations in subsystems of globally entangled quantum states}

\author{Chandan Mahto}
\email{chandanmahto00716@iisertvm.ac.in}
\affiliation{School of Physics, IISER Thiruvananthapuram, Kerala, India 695551}

\author{Vijay Pathak}
\email{vijayp@iisertvm.ac.in}
\affiliation{School of Physics, IISER Thiruvananthapuram, Kerala, India 695551}

\author{Ardra K.~S.}
\affiliation{School of Physics, IISER Thiruvananthapuram, Kerala, India 695551}
\affiliation{Centre for Quantum Dynamics, Griffith University, Gold Coast, QLD 4222, Australia}

\author{Anil Shaji}
\affiliation{School of Physics, IISER Thiruvananthapuram, Kerala, India 695551}

\begin{abstract}
The relation between genuine multipartite entanglement in the pure state of a collection of $N$ qubits and the nonclassical correlations in its two-qubit subsystems is studied. Quantum discord is used as the quantifier of nonclassical correlations in the subsystem while the generalised geometric measure (GGM) [Phys. Rev. A.~{\bfseries 81}, 012308 (2010)] is used to quantify global entanglement in the  $N$-qubit state. While no definite discernible dependence between the two can be found for randomly generated global states, for those with additional structure like weighted graph states we find that local discord is indicative of global multipartite entanglement. Global states that admit efficient classical descriptions like stabilizer states furnish an exception in which despite multipartite entanglement, nonclassical correlation is absent in two qubit subsystems. We discuss these results in the context of mixed state quantum computation where nonclassical correlation is considered a candidate resource that enables exponential speedup over classical computers. 
\end{abstract}

\maketitle

\section{Introduction}
Pinpointing the resources that allow quantum computers to solve certain problems exponentially faster than the best known classical algorithms remains an open question in quantum information theory. In the special case of pure state quantum computing it is known that multiparty entanglement that grows unboundedly with problems size is an essential quantum computational resource~\cite{Jozsa2011}. Quantum coherence and closely related to it, quantum superposition are also considered potential resources for quantum computing and in-depth investigations into the roles of these quantum phenomena in computational speedup has led to the development of detailed resource theories about them~\cite{Chitambar:2019ko,Quantum_speedup,Coh_mix_resource,Coh_ent_resource,Stabilizer_resource,Coherence_resource,Quantum_resource}. 

While substantial progress has been made in identifying the quantum resources that can give computational advantage in various scenarios involving pure quantum states, the same cannot be said about mixed state quantum information processing. On the one hand there is ample evidence that quantum information processors operating on mixed quantum states can provide exponential speedups over classical algorithms for certain computational tasks~\cite{PhysRevLett.81.5672,Quantum_circuit_mix,speedups}. On the other, readily identifiable quantum resources like entanglement, coherence and superposition need not always be present in these states in meaningful quantities for the quantum advantage to be attributable to them~\cite{Role_of_entanglement,Signature_of_nonclassicality,Qc_without_entanglement,
Experimental_Qc_without_entanglement,Qd_in_qc,DQC1Power}.  Nonclassical correlations other than entanglement that can be present in mixed quantum states are considered to be candidate resources that can explain the power of mixed state quantum computing~\cite{DQC1discord}. These correlations can be quantified in terms of measures like quantum discord~\cite{Qd_zurek,Qd_vedral}, measurement induced disturbance~\cite{Qd_mid} etc.  Conclusive evidence to support this suggestion is however still not available. 

An ensemble interpretation is not always available for the mixed state in an information processor since one often considers individual runs of an algorithm on a single device. A fixed number of subsequent runs come into the picture only for collecting measurement statistics just as in the case of pure state quantum computing.  A single copy of a quantum state can be mixed only if it is a subsystem of a larger quantum system that, without loss of generality, can be in a pure state~\cite{nielsen2002quantum,QcandQi_nielson}. Additionally, for the subsystem to be mixed it should be entangled with the rest of the larger quantum system also. Can the mixed state leverage the computational resources of the larger system under certain conditions thereby explaining the quantum advantage furnished by mixed state algorithms? If so, is there a tell-tale signature of the potential to leverage the global resources that can be identified in the mixed states of the subsystems? We investigate these questions under specific scenarios in this paper. 

We limit our discussion to quantum systems that are made up of several qubits. We consider mixed states of a few qubits among them as representative of those used in mixed state quantum computing. We refer to the quantum state of these few - typically two - qubits as the {\em subsystem state} for brevity. The state of the larger collection of qubits is called the {\em global state}. The global state is  assumed to be pure without loss of generality since purification by adding more qubits into the collection on which it is defined is always possible. In effect we have idealized and simplified the picture of an arbitrary mixed quantum state inside a quantum information processor that can be entangled with a potentially large and unknown portion of the rest of the universe down to the case of a few qubit subsystem of a larger collection that is in a pure global state. We investigate the connection between genuine multipartite entanglement involving all the qubits in the global state and the degree of nonclassical correlations in two qubit subsystems. This addresses the second question posed at the end of the previous paragraph directly within the limitations of our simplified model and surprisingly it also partially addresses the first one. 

The structure of this Paper is as follows. In the next section we discuss the measure of multipartite entanglement in the global state and the measure of nonclassical correlations in the subsystem states that we use. Section III focuses on random states of $N$ qubit global systems and their two qubit subsystems. In Section IV we narrow down the discussion to global states that are also resource states for measurement based quantum computation, motivating this case and looking at the nonclassical correlations in subsystems of such states. The subsequent section is on $N$ qubit global state with a specific, all-to-all, multipartite, entanglement structure. A brief discussion of our results and our conclusions are included in Section VI. 

\section{GGM, Discord and Concurrence \label{defs}}

Starting from a global state of $N$-qubits, we are interested in connecting the genuine $N$ party entanglement in it with the nonclassical correlations in two qubit subsystems. Suppose we have only $N-k$ qubit entanglement in the global state. Since we have assumed that the global state is pure, it must then be a product state of two subsystems with $N-k$ and $k$ qubits respectively of the form $\rho_{1}^{N-k} \otimes \rho_{2}^{k}$. If we now take a two qubit subsystem of the global state then two distinct scenarios need to be considered. The first in which the subsystem is either part of $\rho_{1}$ or it is part of $\rho_{2}$. In this case, we can limit the global system to being either $\rho_{1}$ or $\rho_{2}$ as the case may be and the remaining part can safely be ignored. In the second scenario, one of the qubits is part of $\rho_{1}$ and the other is part of $\rho_{2}$. Then there cannot be any nonclassical correlations between the two by assumption since the reduced state of the two qubits will also be a product state and the case is not of interest to us. The same argument can be extended to the case where the $N$ qubits can be split up into multiple subsystems that are not entangled with each other. For this reason we restrict our discussion to $N$ qubit global states that have genuine $N$-party entanglement. 

The Generalised Geometric Measure (GGM) of multipartite entanglement is used as the quantifier for the entanglement in the global state. The starting point for this measure is the geometric measure of entanglement~\cite{Shimony95a,Barnum:2001ea,Wei:2003de} defined for a given pure state $|\psi_{N} \rangle$ on $N$ qubits as 
\begin{equation}
	\label{eq:ggm1}
	{\mathcal G}(|\psi_N\rangle) = 1 - \Lambda_{\rm max}^{2}, \; {\rm where} \; \Lambda_{\rm max} = \max_{|\phi_{N} \rangle} |\langle \phi_{N} | \psi_{N} \rangle|
\end{equation}
 with the maximisation over all $N$ qubit pure product states of the form $|\phi_{N} \rangle = \otimes_{i=1}^{N} |\phi_{i} \rangle$. In~\cite{ggm} this idea was generalised to the GGM by extending the maximisation in Eq.~(\ref{eq:ggm1}) to one over all states $|\phi_{N}\rangle$ that are {\em not} genuinely $N$-party entangled. The measure has the advantage that it is computable for pure states in terms of the Schmidt coefficients across different partitions of the $N$-qubit state as
 \begin{equation}
 	\label{eq:ggm2}
	{\mathcal G}(|\psi_N\rangle) = 1 - \max \big\{ \lambda^{2}_{N-k;k}, \; \forall \; k \big\}.
\end{equation}
Here $\lambda_{N-k;k}$ represents the coefficients appearing in the Schmidt decomposition of $|\psi_{N} \rangle$ with respect to bi-partitions containing $N-k$ and $k$ qubits respectively. The maximisation is not just over the size $k$ of the bi-partitions but also over all possible bi-partitions corresponding to each $k = 1, 2, \ldots \lfloor N/2 \rfloor$. In practice, the GGM computation is simplified further by symmetries in the global state that allows one to consider reduced sets of bi-partitions for each $k$. It turns out that in many of the cases we consider, one typical representative from the $k=1$ case is all that is required to compute the measure. 

We use quantum discord~\cite{Qd_zurek,Qd_vedral} to quantify the nonclassical correlations in two qubit subsystems of the global state. There are several choices of measures of nonclassical correlations but discord is one of the most parsimonious among entropy based measures~\cite{Entropic_measure_nonclassical}, motivating our choice. It is also one of the more widely investigated measurers of such correlations~\cite{Qd_allies,Qd_measures}, particularly in the context of identifying enabling resources for mixed state quantum computation~\cite{Qd_measures,Animeshthesis,Source_speedup}. Since we are dealing with the states of two qubits only, the minimisation over all possible projective measurements on one of the subsystems that appear in the definition of quantum discord can always be done numerically, in those cases where an analytic result is not available.   

Multiparty entanglement of the global state means that bipartite entanglement in the subsystems are typically absent or small due to the monogamy property of entanglement~\cite{PhysRevA.61.052306,Monogamy_entanglement,Horodecki}. We verify that the bipartite entanglement in the subsystems are small by computing the concurrence in the two-qubit subsystem states. A low value for concurrence coupled with a high value for quantum discord allows one to pinpoint the correlations in the subsystem states as nonclassical ones other than entanglement. It also points to the genuine multiparty character of the entanglement in the global state. 

\section{Random pure global states}

\begin{figure}[!tb]
	\resizebox{8.5 cm}{16.5cm}{\includegraphics{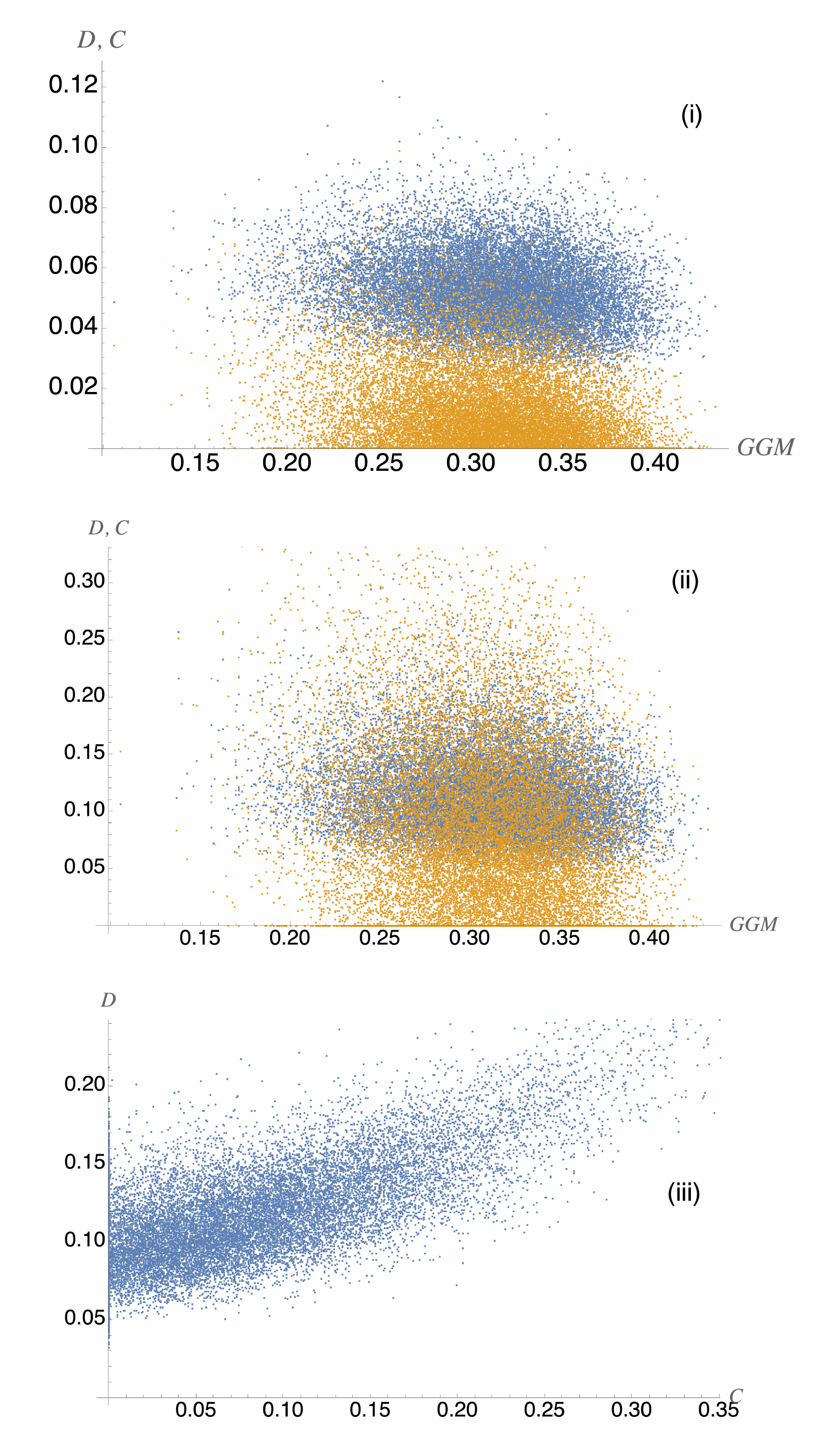}}
	\caption{(Colour Online) Average values of the discord (blue) and concurrence (yellow) of two qubit subsystems of randomly generated 5 qubit states are plotted against the GGM of corresponding global state in (i). In (ii) the maximum values of discord and concurrence are plotted instead of the averages and in (iii) the maximum discord is plotted against the maximum concurrence.  The values shown correspond to 22,000 randomly generated 5 qubit pure states.\label{fig1}}
\end{figure}

It is known that almost all multipartite pure states are entangled~\cite{zyczkowski_volume_1998,lockhart_low-rank_2002} and it is also known that almost all mixed states of such systems have nonclassical correlations~\cite{ferraro_almost_2010}. In the context of the question we are considering, it is therefore natural to explore if there is a direct, quantitative relationship between the multipartite entanglement of randomly chosen global states and the quantum discord in its two qubit subsystems. We generated random $N$ qubit pure states numerically, computed the GGM of these states as well as the average and maximum discord across all possible two qubit subsystems. The average and maximum values of concurrence across all the two qubit subsystems of each randomly generated global state were also computed. We were able to do this optimally for the $N=5$ case. Both the computation of GGM and the computation of discord are numerically challenging and in addition the number of possible two qubit subsystems also rise with $N$.  The results of these computations are shown in Fig.~\ref{fig1}, where we have looked at 22000 randomly generated 5 qubit pure states. We see that there is no discernible definite relationship between the GGM of the global state and the quantum discord in its two qubit subsystems. 

In Fig.~\ref{fig1}-(i) averages of the discord and concurrence taken over all two qubit subsystems of the five qubit state are plotted against the GGM of the global state. We note that almost all the states generated have moderate to high values of GGM as expected and very few states have their GGM lower than 0.2. The average discord in the two qubit subsystems is also significant given that the maximum possible value of two qubit discord is 1. The average concurrence in the two qubit subsystems is significantly lower than the average discord and reflects the expected monogamy properties of quantum entanglement. This also indicates that there is significant amount of nonclassical correlations other than entanglement present in the subsystem states. The inverse relationship between two qubit entanglement and the GGM is showing through, albeit much weakly, in the correlation between discord and GGM. This is further supported by Fig.~\ref{fig1}-(iii) where the maximum discord is plotted against the maximum concurrence. We see that the value of discord is effectively lower bounded by concurrence since the nonclassical correlations quantified by discord includes entanglement also. In Fig.\ref{fig1}-(ii), the maximum values of discord and concurrence respectively taken over all possible two qubit subsystems are shown against the GGM of the global state. This plot also does not reveal any definite relationship between the GGM and nonclassical correlations and follows the same general trends as in Fig.~\ref{fig1}-(i). 

\begin{figure}[!htb]
 	\resizebox{8.5 cm}{6 cm}{\includegraphics{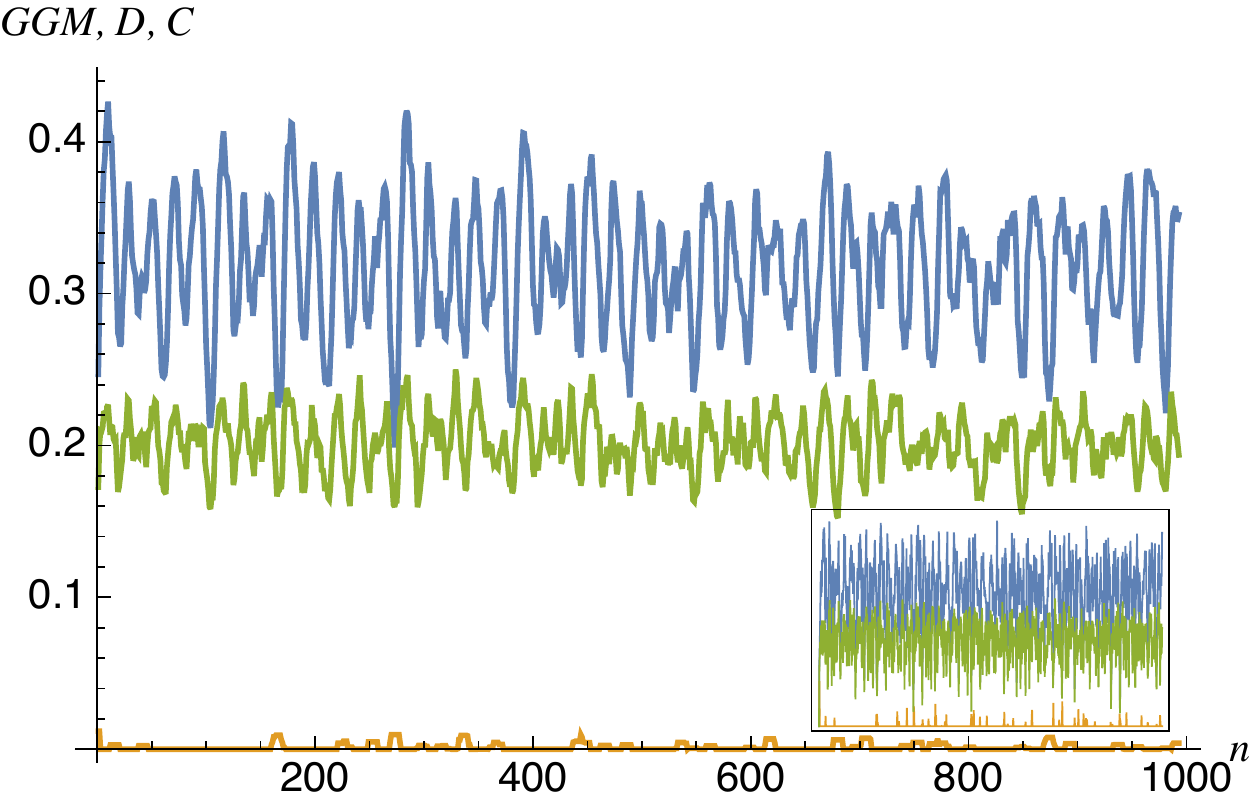}}
	\caption{(Colour online) The GGM of the global pure state of 8 qubits (blue line on top) is computed in addition to the quantum discord in an arbitrarily chosen two qubit subsystem (green line below the blue one) and the concurrence between the two qubits (yellow line close to the $x$-axis) for the kicked quantum top model studied in~\cite{Madhok:2015cs}. The initial state of the qubits as well as the parameters of the kicked top Hamiltonian are chosen identical to the case presented as Fig.~3 (d) in~\cite{Madhok:2015cs}. Only the GGM is computed additionally here. In the main plot, all three quantities have been smoothed out by averaging the value at each point with 4 points each to its left and right to see the trends clearly. The actual data without averaging is shown in the inset.  From the smoothed-out plot, we see that the subsystem discord does follow the global multi-partite entanglement very closely while the lack of any appreciable concurrence between the subsystem qubits indicate that the entanglement present in the global system is genuinely multi-partite in nature. \label{fig2} }. 
 \end{figure}

The lack of a direct and obvious relationship between global multipartite entanglement and nonclassical correlations in the subsystems indicates that the connection between the two may be more subtle. Concluding that such a connection cannot exist is not justified either because for certain classes of states, subsystem discord has been observed to be related to global entanglement. For a multi-qubit system modelled as a quantum kicked top, it is known that as the overall system transitions to a regime where its classical analogue is chaotic, the time averaged discord across any two qubits goes up irrespective of the initial state of the system.  It is also noted that the behaviour of the discord is opposite to that of the two qubit entanglement as quantified by the concurrence~\cite{Madhok:2015cs}.  While the global entanglement is not explicitly computed in this case, it is known that when the kicked top system enters the classically chaotic regime, spreading of entanglement across the system follows~\cite{Wang:2004bz}. In particular, when the classical system is chaotic, irrespective of the initial state chosen, the quantum analogue is expected to have high multi-partite entanglement at all times during the course of its evolution under the kicked top Hamiltonian. We verified this by computing the GGM of the global system in addition to the two qubit subsystem discord and concurrence choosing the system parameters identical to figure 3 (d) of~\cite{Madhok:2015cs}. The results of this computation are shown in Fig.~\ref{fig2}. We see that in the case of $N$ qubits that are initialised in a spin-coherent state and subsequently evolve according to the kicked quantum top hamiltonian, the GGM is consistently high when the equivalent classical system shows fully chaotic behaviour. More importantly, for the question we are considering, we see that the quantum discord in an arbitrarily chosen two-qubit subsystem follows the GGM closely. 

One may, in the present context, treat the presence of chaos in the classical analogue of the kicked top as a witness for multipartite entanglement in the quantum case. In this sense a clear connection between genuine multi-party entanglement in the global state and subsystem discord is seen in the various other cases considered in~\cite{Madhok:2015cs} as well. It is worth noting here that in~\cite{Madhok:2015cs} the initial states chosen for the $N$ qubits are spin coherent states in order to facilitate a comparative study between chaos related features exhibited by the classical version of the kicked top and the behaviour of the corresponding quantum wave function. The spin coherent state is one endowed with more structure than the random states we considered earlier. In particular it has a definite value of the total angular momentum $\vec{J}$ that is preserved by the time evolution under the kicked top Hamiltonian. 

Taking a cue from the kicked quantum top example, we turn our attention towards multi-qubit quantum states endowed with additional structure and properties and ask whether relationships between global entanglement and local discord exist in such cases. A particularly interesting family of states for which genuine multiparty entanglement is assured by construction are the resource states for one way quantum computers~\cite{Raussendorf:2001js} and measurement based quantum computing (MBQC) in general~\cite{briegel_measurement-based_2009,Gross:2007is}. Since MBQC involves performing local measurements on  globally entangled states we expect a detailed study of the relationship between global entanglement and local discord in MBQC resource states to yield insight into the broader question of whether quantum resources of the global state can play a role in enabling exponential speedups in mixed state quantum computation. 

\section{MBQC resource states \label{sec4}}

The paradigm of measurement based quantum computation is recognised as one that is particularly suited for understanding the role of entanglement in quantum information processing. The prototypical model of one way quantum computing~\cite{Raussendorf:2001js,raussendorf_measurement-based_2003,briegel_measurement-based_2009} implements the computation using a resource state that is a multi-qubit entangled state defined on an underlying lattice or graph. Starting from this entangled resource, all further operations are non-entangling ones so that one can zero-in on the initial state exclusively while analysing the role of entanglement in any quantum algorithm implemented as a one-way quantum computation. Graph states on which these computational schemes are implemented have the additional advantage that its structure allows the calculation of reduced density matrices, GGM etc.~analytically. We start with a brief recap of graph states and how to compute the quantities of interest and then move on to specific examples exploring the connection between global entanglement and local discord. 

\subsection{Graph states and their subsystems \label{graph}}

A set of vertices, $V$, together with a collection of edges, $E$, between them form a graph. We limit our discussion to graphs that are undirected, finite and simple (without loops or multiple edges between a pair of vertices). A graph state is constructed by placing qubits on all the vertices and coupling together each pair of qubits that are connected by an edge of the graph through a two qubit unitary transformation $U_{ij}$. The qubits placed on the vertices are typically initialised in the state $|+\rangle = (|0\rangle + |1\rangle)/\sqrt{2}$ and the graph state on $N$ vertices is
\begin{equation}
	\label{graph1}
	|G\rangle = \prod_{(k,l) \in E} U_{kl} |+\rangle^{\otimes N}.
\end{equation}
For simple undirected graphs all two qubit unitaries $U_{kl}$ commute with one another and $U_{kl} = U_{lk}$. Typically MBQC resources states are constructed by applying identical unitaries on all pairs of linked qubits and very often the unitary that is chosen is the {\em Controlled-Z} (CZ) gate that has the representation diag$(1,1,1-1)$ in the computational basis. Fixing the unitary transformation in Eq.~(\ref{graph1}) leads to a single global state of the $N$ qubits. To obtain a one parameter family of  states for which we can explore the nonclassical correlations in their subsystems, we generalise the CZ gate to the  {\em controlled-$\varphi$} ($C_{\varphi}$) operation. The $C_{\varphi}$ gate has the following matrix representation in the computational basis:
\begin{equation}
	\label{cphi}
	C_{\varphi} = \left(  \begin{array}{cccc}
	1 & 0 & 0 & 0 \\ 0 & 1 & 0 & 0 \\ 0 & 0 & 1 & 0 \\ 0 & 0 & 0 & e^{i\varphi}
	\end{array}
\right)
\end{equation}

The  general case wherein each of the unitaries generating the links are controlled phases gates with distinct phases, $U_{kl} = {\rm diag}(1,1,1, e^{i \varphi_{kl} })$, corresponds to a {\em weighted} graph state.  The weights $\varphi_{kl}$ defines the state and the family of states we first look at is the special case corresponding to all the weights being equal to $\varphi$. 
 
 The reduced density matrices for subsystems of weighted graph states can be obtained analytically~\cite{Hartmann:2007db}. Two approaches for constructing the reduced density matrices are described in~\cite{Hartmann:2007db}, of which the one based on a generalisation of projected entangled pair states (PEPS) leads to a pictorial language suited for our discussion. To briefly recap this approach, in lieu of every one of the $N$ qubits in the weighted graph state, $N-1$ virtual qubits are considered. If we focus on an original qubit with label $k$ from the graph state, now represented as $N-1$ qubits, then each one of these virtual qubits are connected to the $N-1$ remaining qubits respectively from the graph.  The link between each virtual qubit $k_{l}$ belonging to the $k^{\rm th}$ original qubit and the virtual qubit $l_{k}$ belonging to the $l^{\rm th}$ original qubit of the graph state is determined by the weight $\varphi_{kl}$ between the two original qubits of the graph. If the two original qubits are not linked to each other then $\varphi_{kl} = 0$ and $U_{kl} = \openone_{kl}$. Specifically, the virtual qubits $k_{l}$ and $l_{k}$ form a generalised entangled pair, $|\varphi_{k_{l} l_{k}} \rangle = U_{kl} |+\rangle_{k_{l}} |+\rangle_{l_{k}}$. The entire graph state of $N$ qubits now gets expanded into $N(N-1)/2$ pairs of qubits with no links across any of the pairs and this state can be written as $|\tilde{\Psi} \rangle = \otimes_{k < l} |\varphi_{k_{l} l_{k}} \rangle$. The graph state $|G\rangle$ is recovered from the PEPS state $|\tilde{\Psi} \rangle$ by projecting at each site $k$ with $P_{k} =|0_{k} \rangle \langle \mathbf{0}_{\bar{k}} | + |1_{k} \rangle \langle \mathbf{1}_{\bar{k}} |$, where $|\mathbf{0}_{\bar{k}} \rangle$ is the $N-1$ (virtual) qubit state $|0_{1} 0_{2} \ldots 0_{k-1} 0_{k+1} \ldots 0_{N} \rangle$ with $|\mathbf{1}_{\bar{k}} \rangle$ similarly defined.  
 
\begin{figure}[!htb]
 	\resizebox{8.5cm}{6.5cm}{\includegraphics{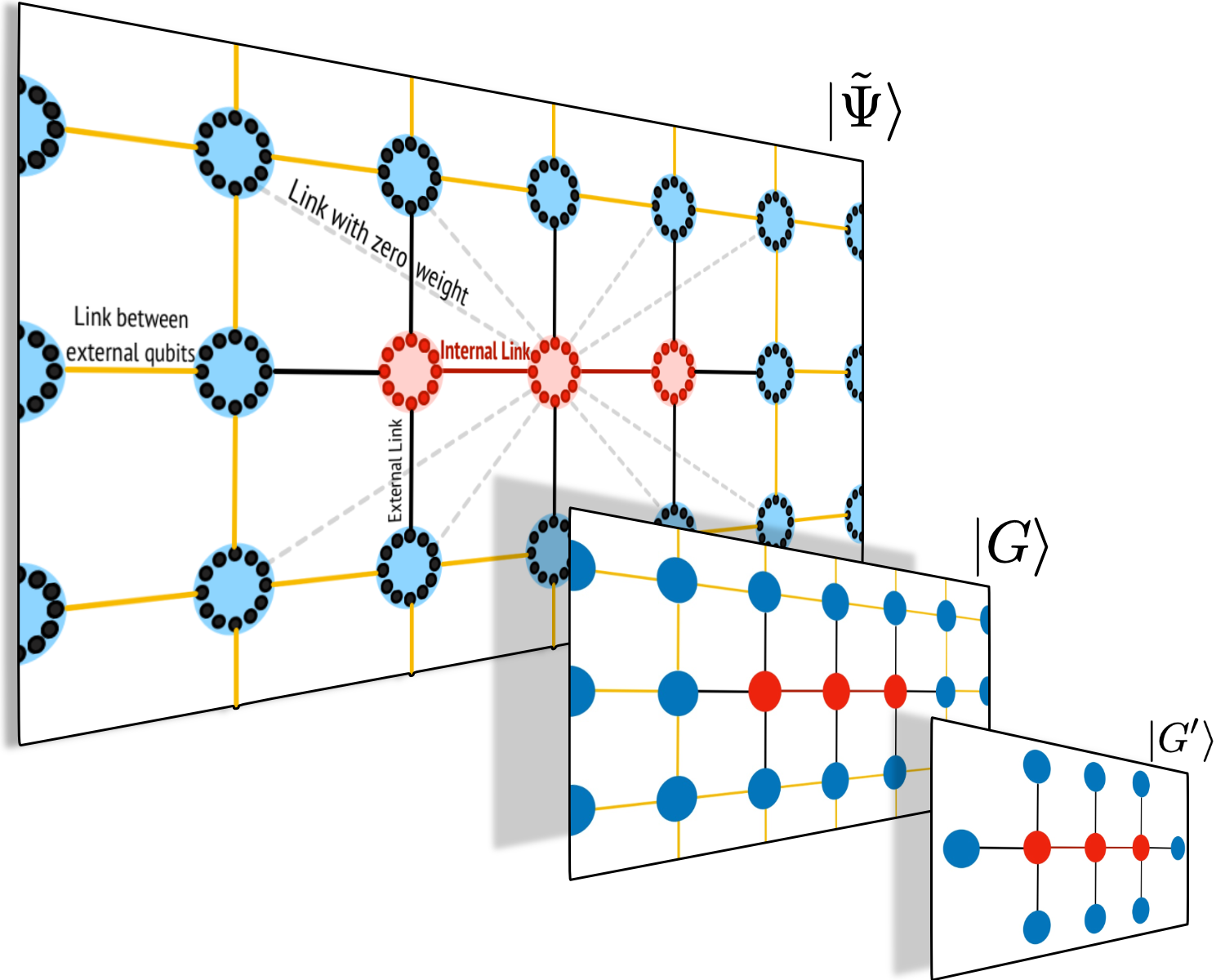}}
	\caption{The panel on the back represents $|\tilde{\Psi}\rangle$ in which each qubit of the graph state has been replaced with $N-1$ virtual qubits. The qubits in red represent the subsystem of interest to us (set $A$) while those in blue are the remaining ones (set $B$). The yellow links are those between qubits in set $B$ while the red links are internal links between qubits in set $A$. The links in black are the ones between qubits in set $A$ and those in $B$. The grey, dashed links represent those links that appear in the PEPS construction with zero weight and these are shown in the figure for only one representative qubit for the sake of clarity. Notice that in the back panel no virtual qubit is linked to more than one other virtual qubit so that we have a collection of independent entangled pairs. The middle panel represents the graph state that is obtained by projecting down $|\tilde{\Psi} \rangle$ with $P_{k}$. The last panel shows the qubits and the links we have to eventually consider if we are to write down the reduced density matrix for the red qubits alone that are in set $A$. Note that the links connecting only blue qubits can now be ignored. \label{figPEPS}}
 \end{figure}

 The PEPS construction ensures that all the $U_{kl}$ are mutually commuting since the unitary operations performed are all on independent pairs of qubits. It also shows that if one is interested only in the reduced state of a subset, $A$, of qubits then all the unitaries that act exclusively between elements of the set $B$ of remaining qubits are irrelevant and they can all be taken to be equal to the identity operator. In other words, after projecting down $|\tilde{\Psi} \rangle$ at each site with $P_{k}$ to obtain the $N$ qubit graph state $|G \rangle$, one can consider a simpler state $|G'\rangle$ in which all links between qubits belonging to $B$ have been removed and write the reduced state of the qubits in set $A$ as
 \[ \rho_{A} = \sum_{kk' \in A} U_{kk'} {\rm Tr}_{B} \big[ |G'\rangle \langle G'| \big] U_{kk'}^{\dagger}.\]
 In the equation above, the unitaries that act among the qubits in $A$ are taken outside the trace over the qubits in $B$ and in the state $|G'\rangle = \prod_{k \in A, l \in B} U_{kl} |+\rangle^{N}$, only the links that connect qubits in set $A$ with those in set $B$ are considered.  The PEPS construction is summarised pictorially in Fig.~\ref{figPEPS} for the case where the graph is a square lattice. In order to evaluate ${\rm Tr}_{B} \big[ |G'\rangle \langle G'| \big]$ we go back to the PEPS picture for only those qubits shown in the third panel of Fig.~\ref{figPEPS} and consider only the links between $A$ and $B$. Applying the projectors $P_{k}$ only on the virtual qubits associated with the blue coloured sites, the following state is obtained:
 \[ |\Psi''\rangle = \bigotimes_{l \in B} \Big[ \Big( \prod_{k_{j}} U_{k_{j}l} \Big) |+\rangle_{\mathbf k} |+\rangle_{l}\Big]. \]
 Note that $|\Psi''\rangle$ is a simple tensor product over all the qubits belonging to $B$. In order to obtain the reduced state of the virtual qubits in $A$, we can trace out each of the qubits in $B$ one-by-one out of this tensor product. Tracing out the $l^{\rm th}$ qubit, the reduced state $\rho_{A}'(l)$ of each of the virtual qubits in subsystem $A$ is obtained as
 \begin{equation}
  	\label{primitive1}
	\rho_{A}'(l) = \frac{1}{2} \big( |+\rangle_{A} \langle +| + |\varphi_{l} \rangle_{A} \langle \varphi_{l}| \big), 
\end{equation}
 where 
 \begin{equation}
  	\label{primitive2}
  	 |\varphi_{l} \rangle = \bigotimes_{k \in A} \frac{1}{\sqrt{2}} \big( |0\rangle + e^{i\varphi_{kl} } |1\rangle \big). 
\end{equation}
The next step is to apply the projectors $P_{k}$ on all the sites in set $A$. The projection leads to a density operator $\rho_{A}'$ for the subsystem of interest, except for the unitaries within subsystem $A$, that is given by a Hadamard product (element-wise multiplication) of all the density operators $\rho'_{A}(l)$ up to a normalisation factor. The proof of this last statement is given in~\cite{Hartmann:2007db}, but pictorially, with reference to Fig.~\ref{figPEPS} one can understand this as each of the blue qubits connected to the set of red qubits contributing a multiplicative factor to each element of the reduced density matrix of the red qubits. In the computational basis, the matrix elements of $\rho_{A}'$ can be written down concisely (setting $\varphi_{kl}^{0} \equiv 0$) as, 
\begin{equation}
	\label{matrixelelements}
	\langle {\mathbf m} | \rho_{A}' |{\mathbf n} \rangle = \prod_{l \in B} \bigg\{ 1 + \exp \bigg[ i \sum_{k \in A} \big(\varphi_{kl}^{n_{k}} - \varphi_{kl}^{m_{k}} \big) \bigg] \bigg\}.
\end{equation}
The last step in obtaining the reduced density matrix, $\rho_{A}$ for the qubits in set $A$ involves applying the unitaries corresponding to the internal links within the set and normalising the density matrix. 

\subsection{Two qubit subsystem of a graph with identical weights}

\begin{figure}[!htb]
 	\resizebox{8 cm}{5.3cm}{\includegraphics{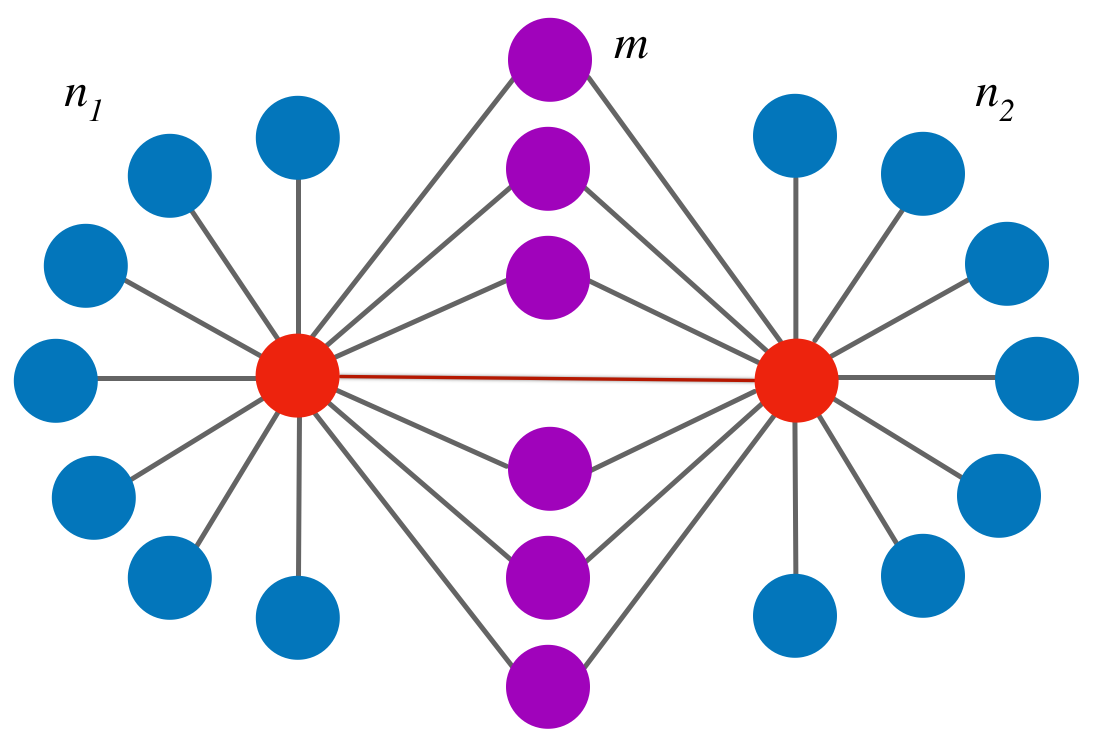}}
	\caption{A generic two qubit subsystem (red dots) of a graph state is shown with all the remaining qubits and links of the graph that are relevant for computing the reduced state of the subsystem. There are $m$ qubits (purple dots) connected to both the qubits in the subsystem of interest while $n_{1}$ qubits (blue dots) are connected independently to one of the subsystem qubits and $n_{2}$ are connected to the second.  \label{primitives}}
 \end{figure}

We are interested specifically in the cases where the set $A$ consists of two qubits only. From here on we assume that all the phases $\varphi_{kl}$ are equal to $\theta$ (all the weights in the weighted graph state are same). The two qubit state corresponding to an arbitrary graph can be obtained using two primitives that are pictorially represented in Fig.~\ref{primitives}. The first primitive is a single qubit connected to $n$ qubits of set $B$ as indicated in the figure by one of the red dots along with all the blue ones connected to it. The reduced density matrix of this qubit is given by
\begin{equation}
	\label{oneqstate}
	\rho_{1}^{n} = \frac{1}{2} \left( \begin{array}{cc} 1 &  e^{in \frac{\theta}{2}} \cos^{n} \frac{\theta}{2} \\  e^{-in\frac{\theta}{2}} \cos^{n} \frac{\theta}{2} & 1 \end{array} \right).
\end{equation}
If the subsystem we are considering consists of two qubits that share no common linked qubit in the set $B$ (no purple qubits), then the two qubit reduced state is just the product state, $\rho_{2} = \rho_{1}^{n_{1}} \otimes \rho_{1}^{n_{2}}$.
 
 To account for qubits in $B$ that are shared between qubits in $A$, denoted by the purple dots in Fig.~\ref{primitives}, we consider the second primitive of two red qubits sharing a single purple qubit. The two-qubit reduced density matrix in this case is
 \[ \rho_{2}' = \frac{1}{4}\left(  \begin{array}{cccc} 
 	1 & e^{i\frac{\theta}{2}} \cos \frac{\theta}{2} &  e^{i\frac{\theta}{2}} \cos \frac{\theta}{2} & e^{i\theta} \cos \theta\\
 	e^{-i\frac{\theta}{2}} \cos \frac{\theta}{2} & 1 & 1 &  e^{i\frac{\theta}{2}} \cos \frac{\theta}{2} \\
	e^{-i\frac{\theta}{2}} \cos \frac{\theta}{2} & 1  & 1 &  e^{i\frac{\theta}{2}} \cos \frac{\theta}{2} \\
 	 e^{-i\theta} \cos \theta &  e^{-i\frac{\theta}{2}} \cos \frac{\theta}{2} &  e^{-i\frac{\theta}{2}} \cos \frac{\theta}{2} & 1
 \end{array} \right). \]
 We can write this density matrix as
 \[ \rho_{2}' = \big( \rho_{1}^{1} \otimes \rho_{1}^{1}  \big) \star \eta_{2}^{1}, \]
 where $\star$ represents element-wise multiplication (Hadamard product) and $\eta_{2}^{m} $ is a `correction' introduced on the simple tensor product on one qubit states by the fact that the external qubit is linked to {\em both} of the qubits we are interested in. We have,
\[ \eta_{2}^{m} =  {\left(  \begin{array}{cccc} 
 	1 & 1 & 1 &P_{m} \\
	1 & 1 & Q_{m} & 1 \\
	1 & Q_{m}& 1 & 1 \\
	P_{m} & 1 & 1 & 1 
  \end{array} \right)}. \]
where
\[ P_{m} =\bigg[ \frac{\cos \theta}{\cos^{2} \theta/2} \bigg]^{m} , \;\; Q_{m} =  \frac{1}{ \cos^{2m} \theta/2 }. \]
  In general, for the two red qubits in Fig.~\ref{primitives} we obtain the reduced density matrix by putting together the two primitives as 
  \begin{equation}
  	\label{eq:gen2q}
	\rho_{2} ({n_{1}, n_{2}, m})=  U_{\theta} \big[ \big( \rho_{1}^{n_{1}+m} \otimes \rho_{1}^{n_{2}+m}  \big) \star \eta_{2}^{m} \big] U_{\theta}^{\dagger},
  \end{equation}
where $U_{\theta}$ accounts for the direct link between the two qubits in set $A$. 
\subsection{GGM for graph states}

The square of the Schmidt coefficients, $\lambda_{N-k;k}^{2}$ that appear in Eq.~(\ref{eq:ggm2}) for computing the GGM are nothing but the eigenvalues of the reduced density matrices of the subsystems that appear when we take all possible bipartitions of the graph state.  If a graph state has a disconnected piece then that subsystem will be in a pure state and $\lambda_{N-k;k}^{2} = 1$. We are however considering only connected graphs. If we further assume that the weights on all the links are identical, we expect the sub-unit (irrespective of the number of qubits in it) that has the least number of external links to be closest to a pure state and therefore have the largest eigenvalue.  For all regular graphs in which the number of links terminating on each qubit is fixed, individual qubits will have the least number of links terminating on them. The number of links terminating on subsystems formed by larger, contiguous groups of qubits will grow according to an area law~\cite{eisert_colloquium_2010}. If the multi-qubit subsystems considered are not contiguous then the number of external links will be even higher. If all the links are equally weighted, it follows that the GGM can then be easily computed by finding the largest eigenvalue of the single qubit reduced density matrix (assuming all qubits have the same number of links and ignoring the edges of the graph where this assumption may not hold). 

A possible exception to the argument above is when the graph is such that all qubits are connected to all others. With $N$ qubits total, each qubit will have $N-1$ links while a $k$ qubit subsystem will have $k(N-k)$ external links. However, if we count the number of distinct external qubits to which the $k$-qubit subsystem is linked to then there are only $N-k$ such distinct links. If we take $k=2$, from Eq.~(\ref{eq:gen2q}) we find that the elements on the counter-diagonal of the two qubit density matrix are changed substantially because of the large number of shared external qubits. This typically leads to increase in purity of the state with corresponding increase in its largest eigenvalue. This makes it likely that the largest eigenvalue of the reduced state of the two qubit subsystem is larger than that of the one qubit subsystem.  We numerically investigated the largest eigenvalues of the reduced states of subsystems of size ranging from 1 to $\lfloor N/2 \rfloor$ of $N$ qubit fully connected graphs for large $N$. We observed that in all the cases, the one qubit sub-system has the highest eigenvalue and the largest eigenvalues decrease with increasing size of the subsystems. For small values of $N$ ($N \leq 4$), the largest eigenvalue of the two qubit reduced is marginally higher than that of the one qubit reduced state for certain values of $\theta$. The take-away message from this analysis is that for all the examples of global graph states we consider that are connected, equally-weighted and invariant under exchange of qubits, the computation of  GGM for the global state effectively reduces to finding the largest eigenvalue of the single qubit reduced density matrix. 

\subsection{Cluster states on a 2D square lattice \label{square}}

We focus initially on resource states defined on a square lattice as per the original conception of the one way quantum computer~\cite{Raussendorf:2001js}. Ignoring the edges of the lattice each qubit is connected to four others and single qubits are also the sub-units with the lowest number of external links. The single qubit reduced density matrix is given by $\rho_{1}^{n}$ from Eq.~\eqref{oneqstate} with $n=4$ and its eigenvalues are $\lambda_{\pm}^{2} = (1 \pm \cos^{4} \theta/2)/2$ leading to the GGM of the cluster state on a 2D square lattice as ${\mathcal G}(\theta) = 1 - \lambda_{+}^{2} $. 

\begin{figure}[!htb]
\resizebox{5.8cm}{8cm}{\includegraphics{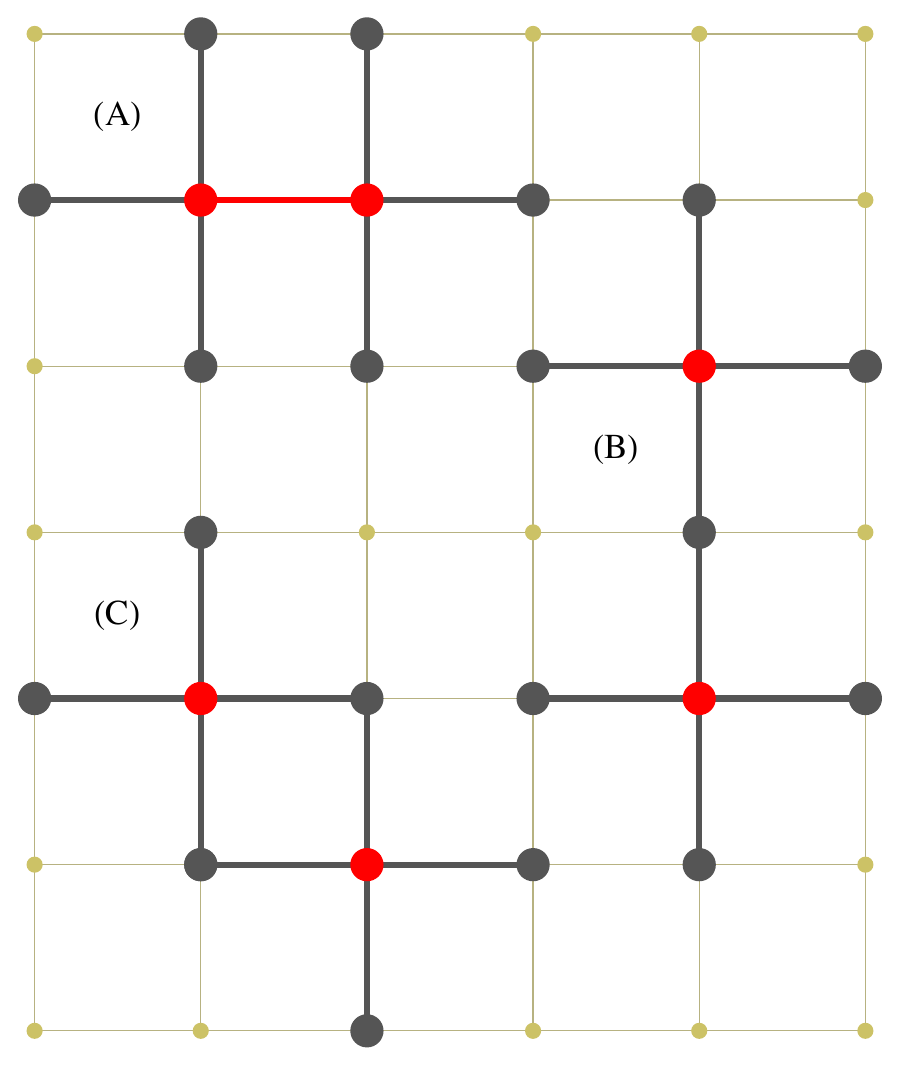}}
	\caption{Three ways in which a pair of qubits that are part of a cluster state on a 2D square lattice can be connected directly or indirectly with each other so that their joint state is not a product state. The two qubits being considered are marked by the red dots. The red line in case (A) denotes a direct link between the two qubits of interest wherein a link corresponds to the two qubit unitary $U(\theta)$ acting on them. In cases (B) and (C) there are no direct connections between the two qubits of interest but they are indirectly linked through shared links to common external qubits. In case (B) there is one such external common linked qubit while there are two such qubits in case (C).    \label{squareconnect}}
\end{figure}

Let us now consider the quantum discord and concurrence of two qubit subsystems of the cluster state.  If the two qubits are completely disconnected from each other in the sense that neither are they connected directly by a lattice edge nor are they indirectly connected through shared edges to common qubits, then their joint state is a product, $\rho_{2} = \rho_{1}^{4} \otimes \rho_{1}^{4}$. This state not entangled and its discord is also zero. We look at pairs of qubits that are connected directly or indirectly and for the 2D square lattice, there are three possible ways in which a pair of qubits can be connected as shown in Fig.~(\ref{squareconnect}). 

\begin{figure}[!htb]
\resizebox{8.5cm}{6cm}{\includegraphics{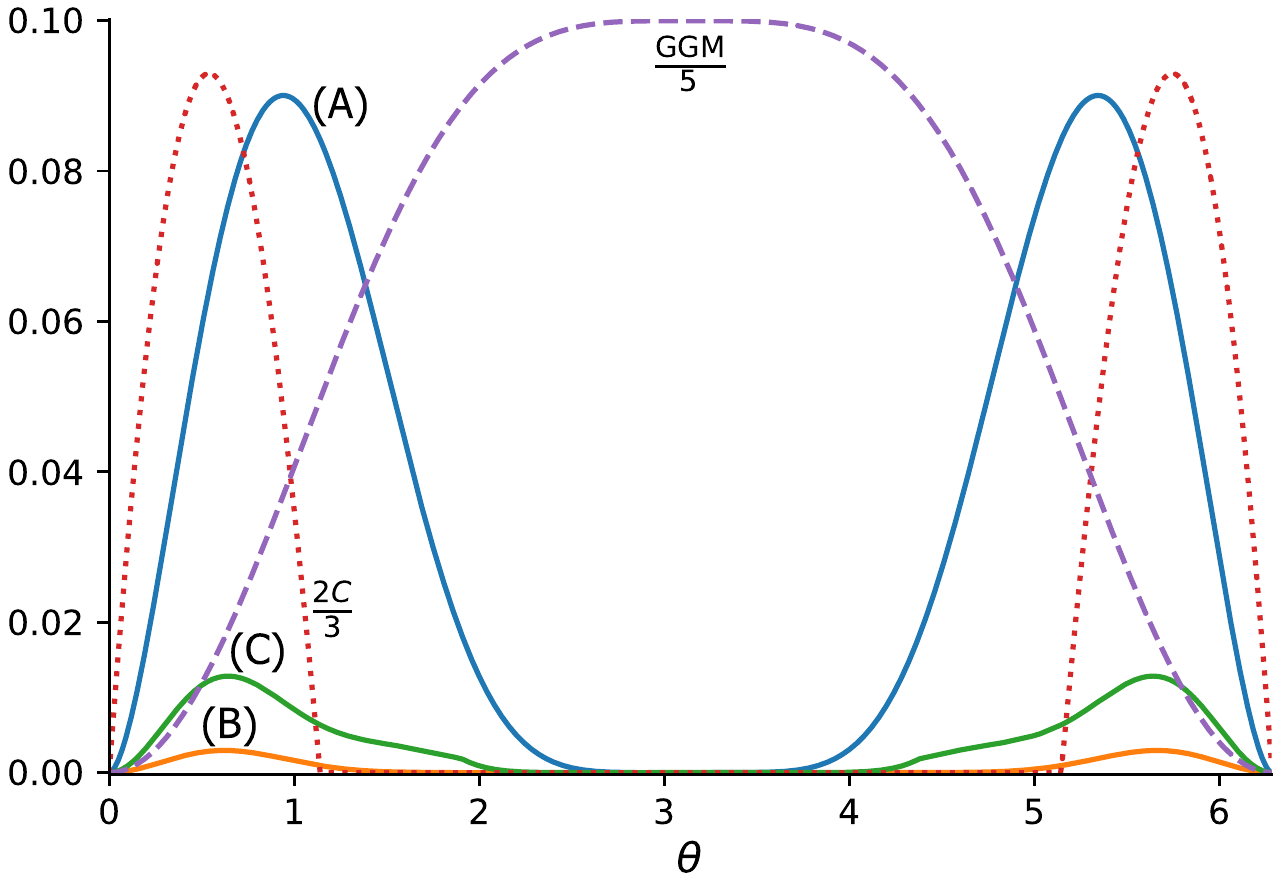}}
	\caption{The solid lines labelled (A), (B) and (C) respectively show the quantum discord of a two qubit subsystem of the cluster state on a 2D square lattice computed for the three ways in which the two qubits can be connected (see Fig~\ref{squareconnect}).  Note that in all three cases, the discord vanishes only at $\theta = \pi$ even if this is not evident from the figure due to the scale used.  The GGM of the cluster state on a 2D square lattice (dashed line) and the the concurrence, $C$, of a pair of directly connected qubits (dotted line) for case (A) are also plotted. Since the form of the graphs are relevant for our discussion the values of GGM and $C$ are scaled down so that all three plots have the same range. \label{2Dsquarediscord}}
\end{figure}

For the three cases of connected pairs of qubits, the respective two qubit reduced density matrices can be obtained using Eq.~\eqref{eq:gen2q} as $\rho_{2}^{(A)} = U_{\theta}^{\vphantom{\dagger}} (\rho_{1}^{3} \otimes \rho_{1}^{3}) U^{\dagger}_{\theta}$,  $\rho_{2}^{(B)} = (\rho_{1}^{4} \otimes \rho_{1}^{4}) \star \eta_{2}^{1}$ and  $\rho_{2}^{(C)} = (\rho_{1}^{4} \otimes \rho_{1}^{4}) \star \eta_{2}^{2}$ respectively.  The quantum discord corresponding to these three cases are obtained using a numerical minimisation of the conditional entropy over all possible projective measurements on one of the two qubits and these are plotted in Fig.~\ref{2Dsquarediscord}.  We note that in case (A) where the two qubits are directly linked by a joint unitary has higher discord than the other two cases. However it is also true that two qubit entanglement as quantified by the concurrence is non-zero only in case (A)  while in the other two cases it is identically zero. The discord in cases (B) and (C) are therefore entirely nonclassical correlations other than entanglement while in case (A) both entanglement and additional nonclassical correlations are present contributing to the higher value of discord. 


A comparison between the GGM of the 2D cluster state and the concurrence with the discord for case (A) is also given in Fig.~\ref{2Dsquarediscord}. In order to focus on the relative behaviour of the three quantities we have scaled the GGM and concurrence such that all three graphs have the same range.  The difference in scales is to the most part because of the fact that we are using different measures that are not directly comparable and this is compensated for by the rescaling done in Fig.~\ref{2Dsquarediscord}. The multipartite entanglement of the global state (GGM) saturates for $\theta$ around $\pi$. As expected from the monogamy of entanglement the concurrence drops to zero when the GGM increases and in the range in which GGM is maximal as well as around it, the concurrence is zero. Interestingly the discord does not go to zero as the GGM saturates and continues to be non-zero except for $\theta = \pi$. The case of $\theta = \pi$ is particularly curious since the corresponding $U_{\theta}$ is the CZ gate. Since CZ is a two qubit gate belonging to the Clifford group, at $\theta = \pi$, it is easy to see that the cluster state is also a stabilizer state that in turn allows an efficient classical description~\cite{aaronson_improved_2004,Stabilizer_resource}. Only for this isolated case wherein the global state can be classical simulated does the nonclassical correlations in the subsystem state completely vanish. We take this as a key point to be discussed in detail after exploring more representative cases.

\subsection{Universal cluster states on other lattices \label{other}}

\begin{figure}[!htb]
\resizebox{3.5cm}{6cm}{\includegraphics{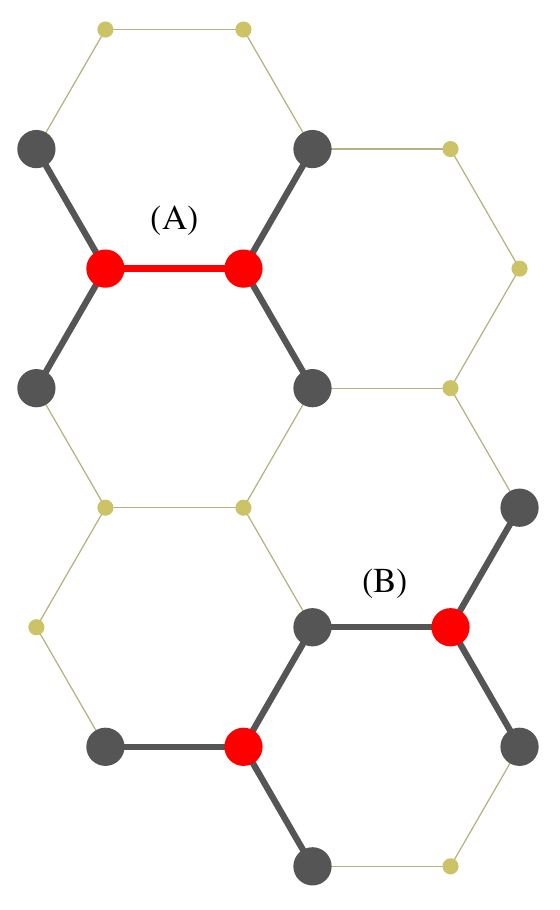}}
\resizebox{4.5cm}{6cm}{\includegraphics{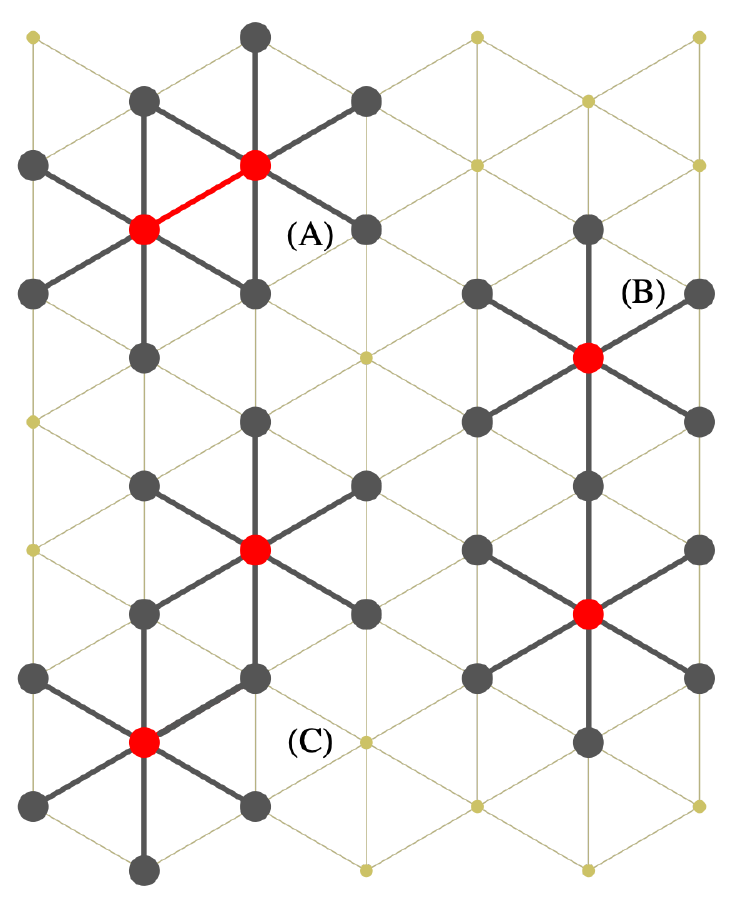}}
	\caption{Two ways in which a pair of qubits (marked in red) that are part of a cluster state on a 2D hexagonal lattice can be connected directly or indirectly with each other are shown on the left. Also shown (right side) are the three ways in which a pair of qubits that are part of a cluster state on a 2D triangular lattice can be connected directly or indirectly with each other.  The red lines in case (A) for both lattices denotes a direct link between the two qubits of interest. In the other cases the two qubits are indirectly linked through shared links to common external qubits.   \label{2Dhex}}
\end{figure}

Universal resources states for measurement based quantum computation are studied in~\cite{VandenNest:2006he} and it is shown that in addition to the square lattice, hexagonal, triangular and Kagome lattices are also universal. Possible ways in which two qubits can be connected without their joint state being a product is shown for the hexagonal lattice and triangular lattices in Fig.~\ref{2Dhex}.

\begin{figure}[!htb] 
\resizebox{8.5cm}{6cm}{\includegraphics{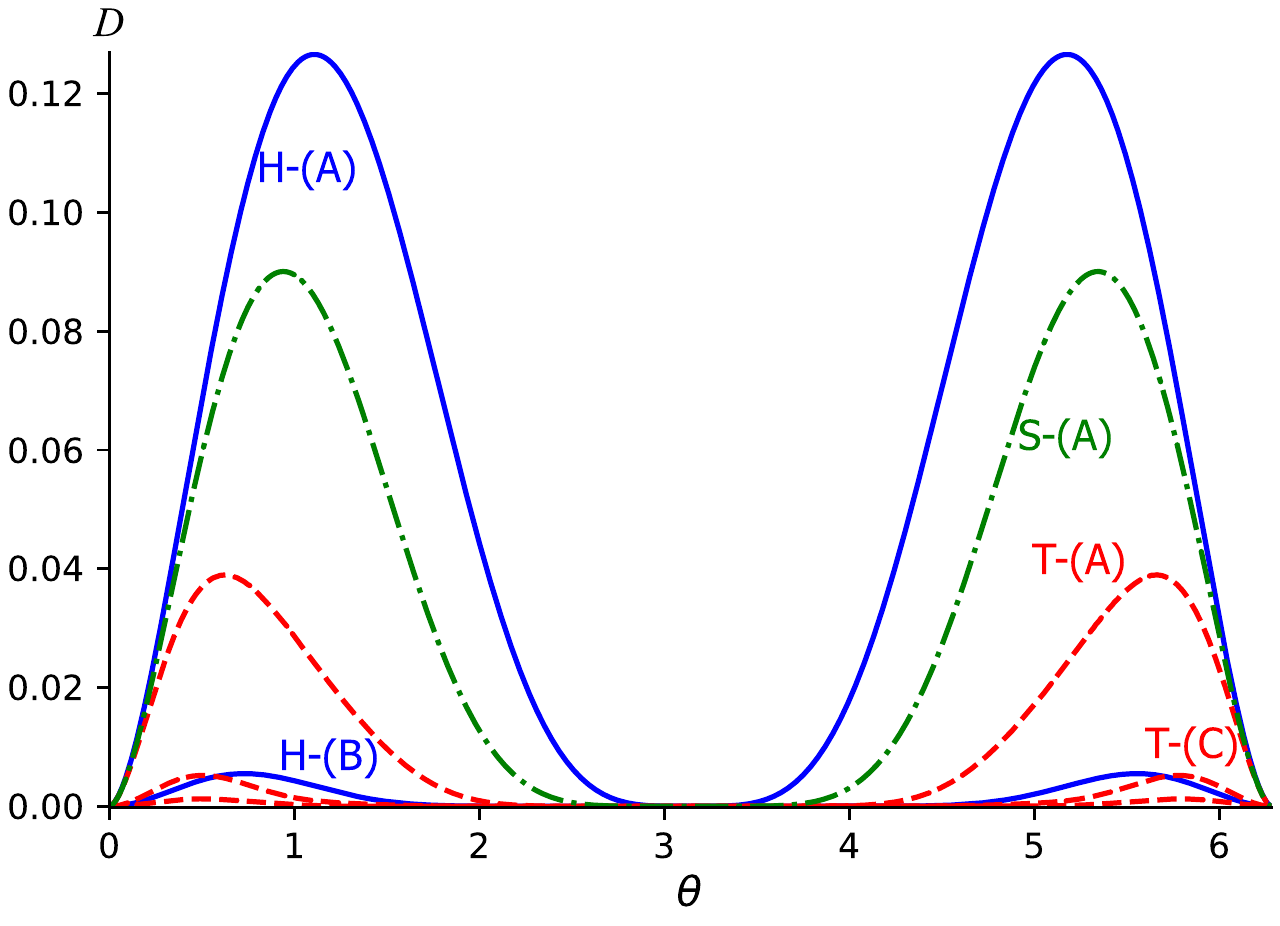}}
	\caption{Discord between the two qubits marked in red in the hexagonal lattice shown in Fig.~\ref{2Dhex} is plotted as a function of $\theta$ using blue (solid) lines. The discord corresponding to the two cases shown for the hexagonal lattice are labelled as H-(A) and H-(B) respectively. Similarly, discord corresponding to the three cases for the triangular lattice are plotted using red (dashed) lines with cases (A) and (C) labelled as T-(A) and T-(C) respectively. The unlabelled red (dashed) plot that is lying very close to the $x$-axis corresponds to case (B) of the triangular lattice. Discord corresponding to case (A) of the square lattice from Fig~\ref{squareconnect} is plotted using green (dot-dashed) line for comparison and labelled as S-(A).   \label{HexTriDisc}}
\end{figure}

The behaviour of the GGM, two-qubit discord and concurrence follows the same pattern as in the case of the square lattice. The discord in all cases goes to zero only when $\theta = \pi$ when the cluster state becomes a stabilizer state. The discord as a function of $\theta$ for the hexagonal and triangular lattices are shown in Fig.~\ref{HexTriDisc}. Also shown in the same figure for comparison is case (A) corresponding to the square lattice. From Fig.~\ref{HexTriDisc} we make the following observations. Comparison of case (A) corresponding to the hexagonal, square and triangular lattices indicate that as the number of external qubits connected to the two qubits we are interested in increases, the discord between the two reduces. Each of the two (red) qubits are connected to two external ones in the hexagonal cases, to three each in the case of the square lattice and five each in the case of the triangular lattice. A comparison of cases (B) and (C) of the square and triangular lattices (See Fig.~\ref{2Dsquarediscord} also) shows that as the number of shared external qubits connected to both the red ones increases the discord also increases and tends to have a higher value when $\theta$ is close to $\pi$. In case (B) for both lattices, the two qubits considered are linked through one shared external qubit each while in case (C) there are two such shared external qubits. These observations prompt us to consider clusters in which all qubits are connected to one another. 

\section{Fully connected clusters \label{fully}}

Consider $N$ qubits in a cluster state with a link between every pair. If we assume that every link is identical and is created by the application of $U_\theta$, then the reduced state of any two qubits in the cluster is given by Eq.~(\ref{eq:gen2q}) as
\begin{equation}
	\label{All2qubit}
	\rho_2(N) = U_\theta^{\vphantom{\dagger}} \tilde{\rho}_2(N) U_{\theta}^{\dagger}, 
\end{equation}
where
\begin{eqnarray*}
	\tilde{\rho}_2(N) & = & \big( \rho_1^{N-2} \otimes \rho_1^{N-2} \big) \star \eta_2^{N-2} \\
	& = &\frac{1}{4} \left( \begin{array}{cccc} 
	1 & A_{\theta/2}^{N-2} & A_{\theta/2}^{N-2} & A_{\theta}^{N-2} \\ 
	\big[ A_{\theta/2}^{N-2} \big]^{*} & 1 & 1 & A_{\theta/2}^{N-2}  \\ 
	\big[ A_{\theta/2}^{N-2} \big]^{*} & 1 & 1 & A_{\theta/2}^{N-2}  \\ 
	\big[ A_{\theta}^{N-2} \big]^* & \big[ A_{\theta/2}^{N-2} \big]^{*} & \big[ A_{\theta/2}^{N-2} \big]^{*} & 1 \end{array} \right),  \nonumber
	\label{All2state}
\end{eqnarray*}
with $A(\varphi) = e^{i\varphi} \cos \varphi$. The corresponding reduced, one qubit density matrix is
\[ \rho_{1}(N) = \frac{1}{2} \left( \begin{array}{cc}  1 & A_{\theta/2}^{N-1} \\ \big[ A_{\theta/2}^{N-1} \big]^{*} & 1 \end{array} \right). \]
The GGM can be calculated from the one qubit reduced density matrix as 
\[ {\mathcal G}(N) = \frac{1}{2} - \frac{1}{2} \bigg| \cos^{(N-1)} \frac{\theta}{2} \bigg|.\]
We see that the for large $N$, the GGM saturates to the value $1/2$ for all values of $\theta$ except for $\theta =0, 2\pi$.

\begin{figure}[!htb] 
\resizebox{8.5cm}{6cm}{\includegraphics{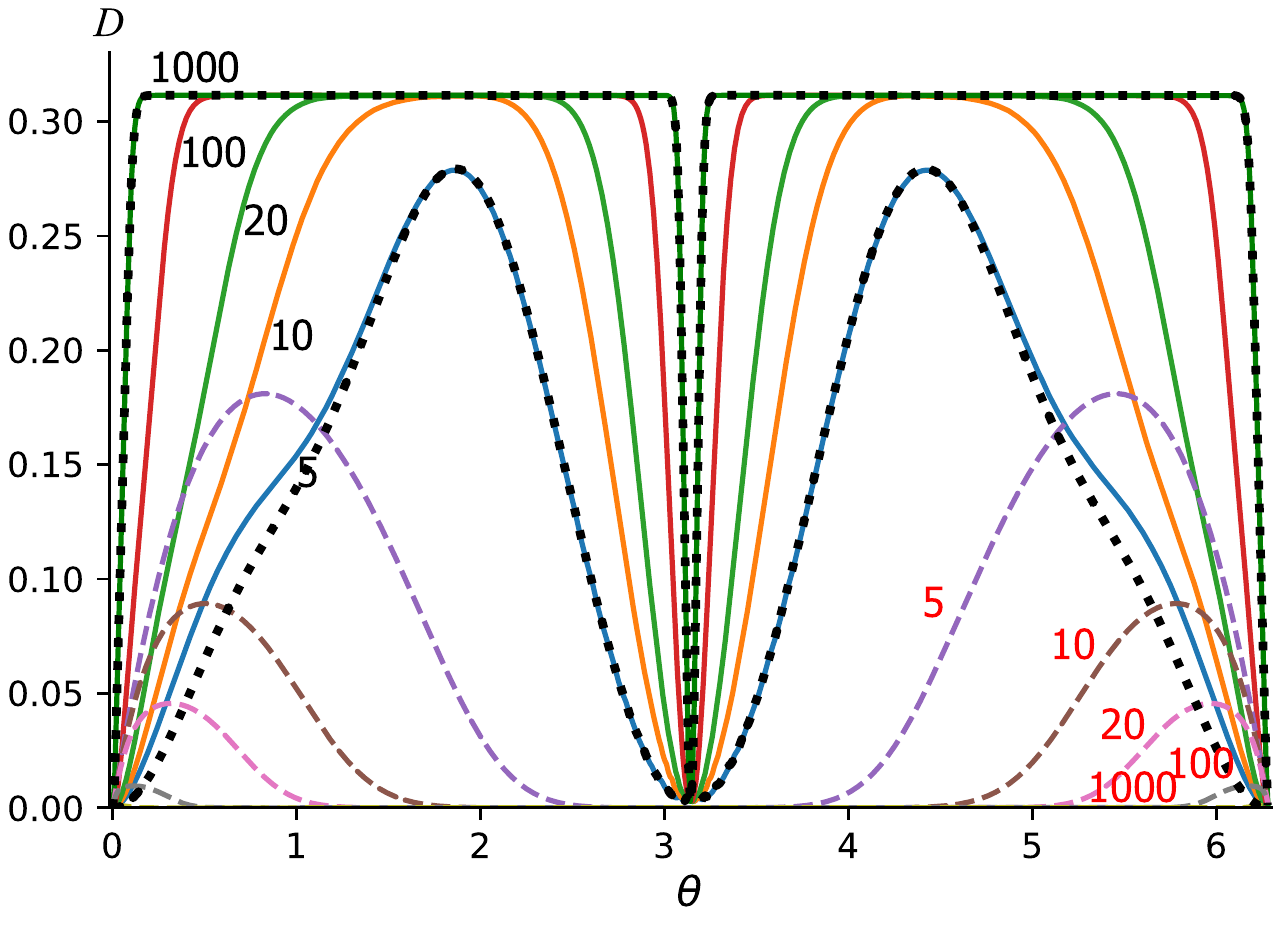}}
	\caption{Discord between a pair of qubits that is part of a fully-connected $N$ qubit cluster state with links formed by $U_{\theta}$ is plotted using solid lines for different values of $N$. As $N$ increases, the discord saturates to its maximum value except at $\theta = \pi$. Also shown as dashed lines (with red labels placed on the right side) are the corresponding values of the concurrence. We see that as $N$ increases, concurrence decreases indicating that for the $N$ qubit, fully connected cluster, the nonclassical correlations in any two qubit system involves no bipartite entanglement as $N$ becomes large. The two black dotted lines show the discord in an $N$ qubit cluster state avoiding the direct link between the two qubits considered. This is shown for $N=5$ and $N=1000$. We see that for large $N$, avoiding the direct link hardly makes any difference to the quantum discord between the two. \label{allQdisc}}
\end{figure}

The two qubit discord computed using the density matrix in Eq.~\eqref{All2qubit} is shown for different values of $N$ in Fig.~\ref{allQdisc}. Also shown in the figure is the two qubit concurrence for different $N$. We see that as $N$ becomes large, the discord saturates to its maximum value for all values of $\theta$ except 0, $\pi$ and $2 \pi$. The concurrence, on the other hand, vanishes completely for large $N$ indicating that entanglement does not contribute to the nonclassical correlations present in two qubit subsystems.  Also shown in the figure as dotted lines is the quantum discord between the two qubits considered if the direct link between them is ignored and their state is taken to be $\tilde{\rho}_2(N)$. We see that as the number of qubits and the number of indirect links connecting the two qubits increases, the relative importance of the direct link in producing discord between the two diminishes. In other words $D \big[ \tilde{\rho}_2(5) \big] \neq D \big[ \rho_2(5) \big]$ while $D \big[ \tilde{\rho}_2(1000) \big] \simeq  D \big[ \rho_2(1000) \big]$.

At $\theta = \pi$, the GGM is still maximum but irrespective of $N$ the two qubit discord goes to zero. As pointed out earlier, at $\theta = \pi$, the $N$ qubit state is also a stabilizer state that allows for an  efficient classical description. This means that for the class of graph states considered here, global entanglement that cannot be classically described is specifically reflected in the subsystems through finite discord between them even if the bipartite entanglement they share is zero. 

\subsection{Fully connected randomly weighted graphs}

\begin{figure}[!htb] 
\resizebox{8.5cm}{6cm}{\includegraphics{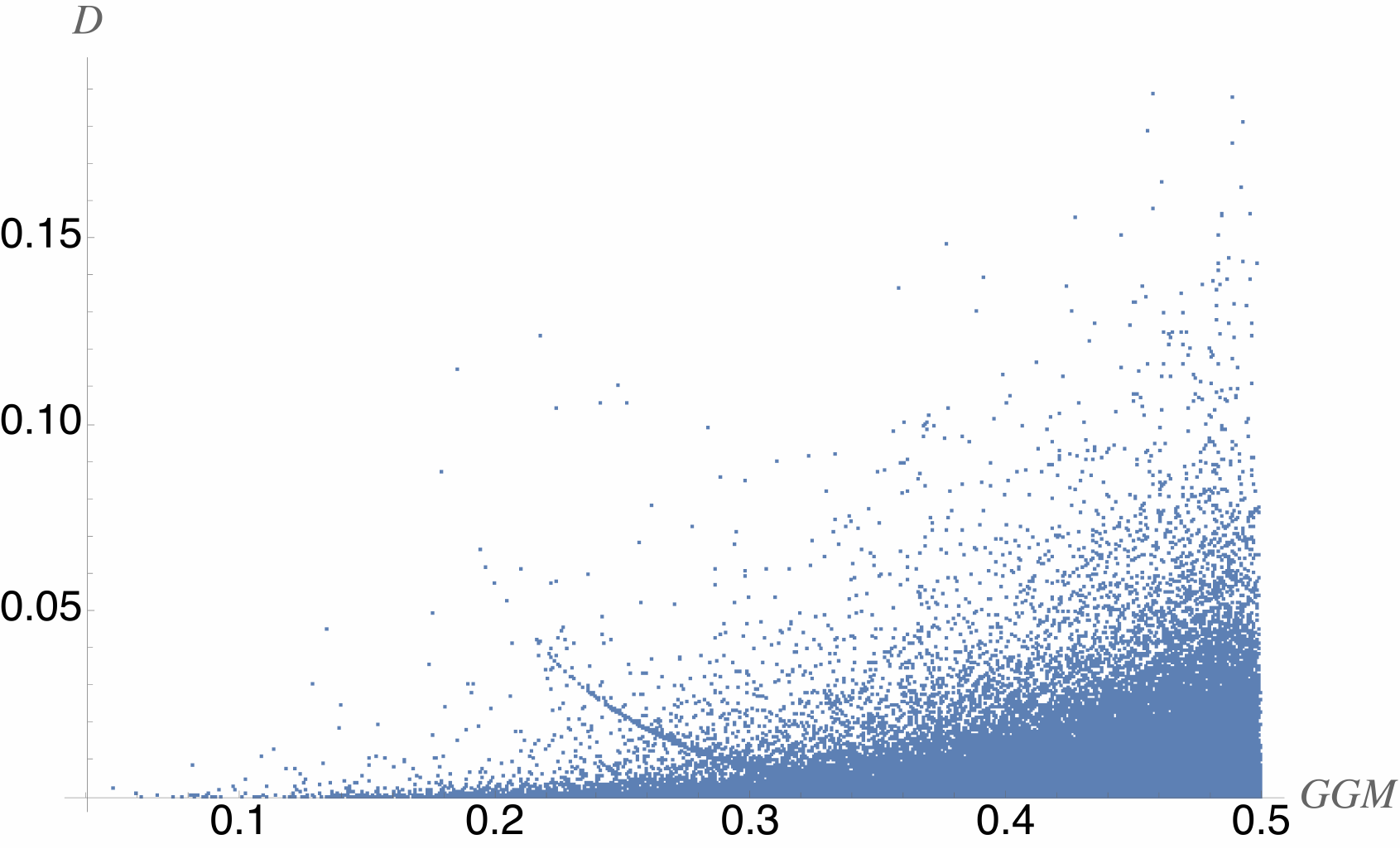}}
	\caption{Discord between qubits 1 and 2 plotted against the GGM for 635,000 instances of fully connected randomly weighted graph states with $N=10$. The link between qubits 1 and 2 is kept the same in all instances and is chosen as the CZ gate ($\theta = \pi$). We see that in contrast to the case in Fig.~\ref{fig1}, there is a definite relationship between the discord and GGM that can be discerned with higher values of discord being associated with higher values of GGM also.  \label{allUrand}}
\end{figure}

We now consider a pair of qubits randomly picked from a collection of $N$ qubits that are all assumed to have interacted with one another. We relax the restriction that all the links between the qubits have to be identical and consider a $N$ qubit, fully connected cluster with random weights (angle $\theta$) chosen for each of the links in the graph. We keep the link between the two qubits we consider constant and choose it to be the CZ gate while varying all the other links randomly. Keeping $N=10$, we generated 635,000 instances of such fully connected graph states and in each case computed the GGM and the discord between two qubits that are labelled as qubits 1 and 2. A scatter plot of the quantum discord versus the GGM is given in Fig.~\ref{allUrand}. 

We see from Fig.~\ref{allUrand} that there is a positive correlation between the GGM of the global state and the quantum discord in a two qubit subsystem. It is instructive to compare Fig.~\ref{allUrand} with Fig.~\ref{fig1} where we have plotted GGM versus discord for global states that are generated by the application of random $N$-qubit unitary transformations acting on an initial product state. The fully connected structure of the $N$ qubit graph states leads to higher values of subsystem discord being associated with larger values of GGM of the global states. In this case again subsystem entanglement is absent and concurrence values were found to be less than $10^{-8}$ for all randomly weighted states that we generated. In the case of the random states we considered in Fig~\ref{fig1},  bipartite entanglement is typically non-zero and accounts for a part of the discord between in the subsystem considered. In the present case, there is hardly any contribution from two qubit entanglement to the nonclassical correlations that exist between them.

\section{Discussion and conclusion \label{conclusion}}

We have established a clear connection between the nonclassical correlations in two-qubit subsystems of $N$-qubit pure states with genuine multipartite entanglement provided the global state has additional structure. We started with two questions about the quantum resources that allow mixed state quantum computers to perform certain tasks exponentially faster than classical computers. We asked if the computational capabilities of the global state that the mixed state in question is part of are reflected in a recognisable manner in the subsystem mixed state. For global states that are also graph states, we find that nonclassical correlations in the subsystem state is indeed indicative of global multipartite entanglement. The interesting exception to this is when the global state is an entangled stabilizer state that admits an efficient classical description. A graph state constructed using CZ gates is a stabilizer state but not all stabilizer states are graph states. However, as shown in~\cite{anders_fast_2006}, every stabilizer state can be mapped to a stabilizer graph state through local stabilizer operations.  Moreover, stabilizer graph states provide a means of efficiently simulating using classical means all other stabilizer states as well as the action of additional stabilizer operations on them. If we take a resource theoretic approach and consider only those quantum states that cannot be simulated efficiently on classical computers as being resources for quantum computation, we conclude that among graph states, such resource states not only have global entanglement but also that their subsystems typically contain nonclassical correlations. This answers the second question we posed in the affirmative for the global states with various kinds of additional structures that we considered.   

The previously studied example of quantum kicked top~\cite{Madhok:2015cs}  also points in the same direction as our analysis of graph states in that globally entangled states are accompanied by local discord. The challenge in the case of the kicked top is in quantifying using a suitable measure the global entanglement signalled by the presence of chaos in the classical limit.  The computation of GGM in the examples we considered were greatly simplified by the observation that in almost all cases, it is sufficient to compute the eigenvalues of the single qubit reduced density matrix. This is not necessarily the case for generic quantum states including the states of the kicked top and calculating the GGM is computationally hard. The availability of a measure of genuine multipartite that is computable relatively easily is the main challenge that must be overcome for extending our results to more general cases. 

It is instructive to imagine a hypothetical mixed state quantum computer in which all $n$ qubits of the input register are part of a global pure graph state of $N$ qubits. We assume that there are links between all $N$ qubits except for the direct links between the $n$ input ones. If the underlying graph state is a stabilizer state then nonclassical correlations shared between any pair of qubits in the input register would be zero. In other words, the input register would be in a concordant state~\cite{cable_exact_2015,eastin_simulating_2010}. Any computational process on the $n$ qubit register can be idealised as a nonClifford unitary that links all the qubits in the input register together. In this picture, the computational process starts introducing non-Clifford gates and links between the $n$ input qubits and thereby also completes the $N$ qubit global pure graph state with nonClifford links and gates. Depending on the structure of the graph, it may very well be that no entanglement is generated between the $n$ qubits in the quantum computer. Only nonclassical correlations other than entanglement may be produced as in the examples  we have seen earlier of the fully connected graphs. The global state however has genuine multipartite entanglement and the introduction of the non-Clifford links means that the Gottesmann-Knill route for describing it efficiently using classical means is no longer available as the the computation progresses on the $n$-qubit subsystem. In this picture, one finds some support for the conjecture that entanglement of a computationally useful type in the global state might be the key resource that can explain the superior performance of mixed state quantum computers relative to classical ones in certain scenarios. This partially addresses the first question we had posed at the outset.

Our investigations into the possible connections between global multipartite entanglement and nonclassical correlations in mixed states of subsystems are still very much only an interesting and insightful starting point since it is limited to only a few families of global states due to limitations in computing quantitative measures of multipartite entanglement. However it gives further support for considering nonclassical correlations as the candidate resource that endows mixed state quantum computers with the ability to perform certain computations exponentially faster than classical ones as in the case of the DQC1 circuit~\cite{DQC1discord}. The results presented here point to an understanding of how these nonclassical correlations in-fact may be signalling a role for the entanglement in an extended state in enabling the computational speedup. 

\section*{Acknowledgements}
A.~S.~acknowledges the support the QuEST program of the Department of Science and Technology through project No. Q113 under Theme 4. V.~P.~acknowledges support from CSIR through Fellowship Grant No. 09/997(0040)/2015 EMR-I. All authors acknowledge the centre for high performance computing of IISER TVM for the use of the HPC cluster, {\em Padmanabha}.

\bibliography{SubsystemDiscord}

\end{document}